\documentclass[]{aa}
\usepackage{amsmath}
\usepackage{graphics}
\usepackage{natbib}
\usepackage{amsmath}
\usepackage{amssymb}

\begin{document}

\title{Detection of small magnetic flux ropes from the third and fourth Parker Solar Probe encounters}
\author{L.-L.~Zhao\inst{1}
\and G.~P.~Zank\inst{1,2}
\and Q.~Hu\inst{1,2}
\and D.~Telloni\inst{3}
\and Y.~Chen\inst{2}
\and L.~Adhikari\inst{1}
\and M.~Nakanotani\inst{1}
\and J.~C.~Kasper\inst{4,5}
\and J.~Huang\inst{4}
\and S.~D.~Bale\inst{6,7,8,9}
\and K.~E.~Korreck\inst{5}
\and A.~W.~Case\inst{5}
\and M.~Stevens\inst{5}
\and J.~W.~Bonnell\inst{7}
\and T.~Dudok de Wit\inst{10}
\and K.~Goetz\inst{11}
\and P.~R.~Harvey\inst{7}
\and R.~J.~MacDowall\inst{12}
\and D.~M.~Malaspina\inst{13}
\and M.~Pulupa\inst{7}
\and D.~E.~Larson\inst{7}
\and R.~Livi\inst{7}
\and P.~Whittlesey\inst{7}
\and K.~G.~Klein\inst{14,15}
\and N.~E.~Raouafi\inst{16}
}

\institute{Center for Space Plasma and Aeronomic Research (CSPAR), The University of Alabama in Huntsville, Huntsville, AL 35805, USA  
\and Department of Space Science, The University of Alabama in Huntsville, Huntsville, AL 35899, USA  
\and National Institute for Astrophysics Astrophysical Observatory of Torino Via Osservatorio 20, 10025 Pino Torinese, Italy  
\and Department of Climate and Space Sciences and Engineering, University of Michigan, Ann Arbor, MI 48109, USA  
\and Smithsonian Astrophysical Observatory, Cambridge, MA 02138 USA  
\and Physics Department, University of California, Berkeley, CA 94720-7300, USA  
\and Space Sciences Laboratory, University of California, Berkeley, CA 94720-7450, USA  
\and The Blackett Laboratory, Imperial College London, London, SW7 2AZ, UK  
\and School of Physics and Astronomy, Queen Mary University of London, London E1 4NS, UK  
\and LPC2E, CNRS and University of Orl\'eans, Orl\'eans, France  
\and School of Physics and Astronomy, University of Minnesota, Minneapolis, MN 55455, USA  
\and Solar System Exploration Division, NASA Goddard Space Flight Center, Greenbelt, MD 20771, USA  
\and Laboratory for Atmospheric and Space Physics, University of Colorado, Boulder, CO 80303, USA  
\and Lunar and Planetary Laboratory, University of Arizona, Tucson, AZ 85721, USA  
\and Department of Planetary Sciences, University of Arizona, Tucson, AZ 85719, USA  
\and Johns Hopkins University Applied Physics Laboratory, Laurel, MD 20723, USA 
}

\abstract{} 
{We systematically search for magnetic flux rope structures in the solar wind to within the closest distance to the Sun of $\sim$0.13 AU, using data from the third and fourth orbits of the Parker Solar Probe.} 
{We extended our previous magnetic helicity-based technique of identifying magnetic flux rope structures. The method was improved upon to incorporate the azimuthal flow, which becomes larger as the spacecraft approaches the Sun.} 
{A total of 21 and 34 magnetic flux ropes are identified during the third (21-day period) and fourth (17-day period) orbits of the Parker Solar Probe, respectively. We provide a statistical analysis of the identified structures, including their relation to the streamer belt and heliospheric current sheet crossing.} 
{} 

\titlerunning{Detection of small magnetic flux ropes}
\authorrunning{L.-L.~Zhao et al.}

\maketitle

\section{Introduction}\label{sec:introduction}
The \textit{Parker Solar Probe} (\textit{PSP}) was launched in August 2018 and completed five orbits around the Sun by August 2020.
During the first three orbits, \textit{PSP} reached a radial distance of $\sim$0.17 AU from the Sun \citep{Fox2016SSR}.
Several curious results were obtained from in situ measurements of the solar wind plasma and magnetic field, including the presence of frequent magnetic switchbacks and surprisingly large rotational flows \citep{Kasper2019, Bale2019}.
After its fourth perihelion, the closest radial distance between \textit{PSP} and the Sun was further reduced to $\sim$0.13 AU, providing an opportunity to study an unexplored regime.

The solar wind is a natural laboratory for studying the physics of turbulent fluctuations \citep[e.g.,][]{Bruno2013}, and the evolution of solar wind turbulence is a major question that is to be addressed by \textit{PSP}.
A common view of solar wind turbulence is based on the nearly incompressible (NI) magnetohydrodynamic (MHD) model.
The NI model suggests that the majority of the turbulent fluctuation energy resides in quasi-2D modes when the plasma beta $< 1$ or $\sim 1$ \citep{Zank1992, Zank1993, Hunana2010, Zank2017}.
It has been suggested that small-scale magnetic flux ropes (SFRs) observed in the solar wind may be indicative of quasi-2D MHD turbulence \citep[e.g.,][]{Greco2009, Zank2018, Zank2020}.
Magnetic flux ropes are characterized by helical magnetic field lines wrapped around an axial magnetic field. They are also called magnetic islands when viewed in 2D.
Properties of SFRs in the solar wind have been studied frequently near 1 AU \citep{Cartwright2010, Kilpua2009, Yu2014, Zheng2018, Hu2018}. They commonly have a duration that lasts from a few minutes to a few hours with a scale size of less than 0.01 AU. 
The statistical analysis of SFRs indicates that they originate from local solar wind turbulence \citep{Hu2018}.
Another possibility is that SFRs originate directly from the Sun and that they are indicative of the connectivity of the solar coronal magnetic field \citep{Borovsky2008}.
Some of the observed SFRs may be related to narrow coronal mass ejections (CMEs), which are expelled from the Sun \citep{Sanchez2017, Rouillard2010a, Rouillard2010b}.
It is worth noting that there is a specific kind of SFR, which is usually observed to be embedded in the sheath of the host CME or interplanetary coronal mass ejection (ICME). They originate from the Sun and have been described as ``ICME-in-sheath." They are usually short and last for a few hours at 1 AU. Inside these SFRs, solar wind parameters are greatly enhanced, especially the magnetic field strength, which is due to the compression of shock and the host ICME \citep{Liu2020ApJL}.

Early observations of magnetic flux ropes relied on visible signatures of magnetic field rotation or the magnetic hodogram \citep[e.g.,][]{Burlaga1981JGR, Lepping1990JGR, Moldwin1995, Khabarova2015}. The Grad-Shafranov (GS) method \citep[e.g.,][]{Sonnerup1996, Hau1999} is useful for the reconstruction of flux rope structures \citep[e.g.,][]{Hu2001GeoRL, Hu2002JGRA, Zheng2018, Hu2018}. In particular, \cite{Liu2008ApJ} verified the flux-rope geometry of CMEs by applying the GS reconstruction method to well-separated multi-spacecraft in situ measurements. 
In a previous study, we developed a technique to systematically identify magnetic flux rope structures using the first orbit measurements of \textit{PSP} \citep{Zhao2020ApJS}. The technique is based on a wavelet analysis \citep{Torrence1998} of the normalized reduced magnetic helicity \citep{Matthaeus1982}.
Magnetic flux rope structures are primarily identified from an enhanced magnetic helicity \citep{Telloni2012ApJ, Telloni2013ApJ}, indicating helical magnetic field lines.
However, a high magnetic helicity is not unique to magnetic flux ropes. Alfv\'en waves or Alfv\'enic structures \citep[e.g.,][]{Alexandrova2006} may also have a high magnetic helicity.
To distinguish magnetic flux ropes from Alfv\'enic structures, the normalized cross helicity and the normalized residual energy are evaluated within the identified structures. Magnetic flux ropes are structures with a normalized cross helicity close to zero and a negative normalized residual energy; by contrast, Alfv\'enic structures have a normalized cross-helicity close to $\pm$1 and null residual energy \citep{Zhao2019ApJb, Zhao2020ApJS}. A comparison of the magnetic helicity-based detection method with the GS reconstruction technique has been presented in \cite{Zhao2019ApJb} and \cite{Chen2020}, and the results show that the two methods are reasonably consistent. 

Observations from the first orbit of \textit{PSP} show that magnetic flux ropes are mostly observed in slow-speed solar wind, while the fast solar wind is dominated by Alfv\'enic structures \citep{Zhao2020ApJS}.
However, there is a caveat to this conclusion as the fast solar wind flows observed by \textit{PSP} are typically highly field-aligned. Based on Taylor's hypothesis, magnetic flux ropes, which are quasi-2D structures with wavevectors that are perpendicular to the mean magnetic field, cannot be observed when the solar wind flow is field-aligned.
In this paper, we extend the analysis to data from the third and fourth orbits of \textit{PSP} which cover a radial distance from $\sim$0.13 to $\sim$0.6 AU.
A commonly made assumption in calculating the reduced magnetic helicity is that the solar wind flow velocity is radial, as in our previous study \citep{Zhao2020ApJS}.
Such a condition is usually warranted at 1 AU or beyond, but as \textit{PSP} continues to approach the Sun, the azimuthal flow velocity becomes more significant \citep{Kasper2019}.
During the first three orbits, the transverse flow speed was as high as $\sim 50$ km/s near perihelia.
In this work, a more general formula for magnetic helicity is utilized, which takes an arbitrary flow direction into account. We discuss the detection technique in Section \ref{sec:technique}.
The results are shown in Section \ref{sec:results}. 
An estimate of the uncertainty is presented in Section \ref{sec:error}.
Section \ref{sec:summary} provides a summary and conclusions.

\section{Magnetic flux ropes detection technique}\label{sec:technique}
\subsection{Evaluation of magnetic helicity}

The evaluation of magnetic helicity with spacecraft data is based on the approach of \citet{Matthaeus1982}. Although the formula in \citet{Matthaeus1982} is given for the case of purely radial velocity, it can be easily extended to an arbitrary flow direction.
We start with the relation
\begin{equation}\label{eq:magnetic}
  \vec{B} = \nabla \times \vec{A},
\end{equation}
where $\vec{B}$ is the magnetic field and $\vec{A}$ is the vector potential. Assuming $\nabla \cdot \vec{A} = 0$ and using the curl of Equation \eqref{eq:magnetic}, we find
\begin{equation}
  \nabla^2 \vec{A} = -\nabla \times \vec{B}.
\end{equation}
In Fourier space, we let $\nabla \to \mathrm{i}\vec{k}$, so that
\begin{equation}
\tilde{\vec{A}}  =  \mathrm{i} \frac{\vec{k} \times \tilde{\vec{B}}}{k^2} \quad\mathrm{or}\quad  
  \tilde{A}_i  =  \frac{\mathrm{i} \epsilon_{ilm} k_l \tilde{B}_m} {k^2},\label{eq:Ai}
\end{equation} 
where the tilde represents Fourier transformed quantities and $\epsilon_{ilm}$ is the antisymmetric tensor.
The magnetic power spectrum matrix $S_{ij}$ is defined as the Fourier transform of the magnetic correlation matrix $\left<B_i(\vec{x})B_j(\vec{x} + \vec{r})\right>$ or in terms of the Fourier transform of the magnetic field components,
\begin{eqnarray}
  S_{ij} &=& \int\left<B_i(\vec{x})B_j(\vec{x}+\vec{r})\right> \mathrm{e}^{-\mathrm{i} \vec{k}\cdot\vec{r}} \mathrm{d}^3 \vec{r} \nonumber \\
         &=& \lim_{V\to\infty} \left<\tilde{B}_i^*(\vec{k})\tilde{B}_j(\vec{k})\right>/V,
\end{eqnarray}
where $V$ is the volume. The asterisk represents the complex conjugate, as does the asterisk below. Similarly, the cross spectrum of $\vec{A}$ and $\vec{B}$ can be defined as
\begin{eqnarray}
  H_{ij} &=& \int\left< A_i(\vec{x})B_j(\vec{x}+\vec{r})\right> \mathrm{e}^{-\mathrm{i} \vec{k}\cdot\vec{r}} \mathrm{d}^3 \vec{r} \nonumber \\  
         &=& \lim_{V\to\infty} \left<\tilde{A}_i^*(\vec{k})\tilde{B}_j(\vec{k})\right>/V.
\end{eqnarray}
Using Equation (\ref{eq:Ai}), we find
\begin{eqnarray}
  H_{ij} &=& \lim_{V\to\infty}\left<-\frac{\mathrm{i}\epsilon_{ilm}k_l\tilde{B}_m^*(\vec{k})}{k^2} \tilde{B}_j(\vec{k})\right>/V \nonumber \\
         &=& -\frac{\mathrm{i}\epsilon_{ilm}k_l}{k^2} S_{mj}(\vec{k}).
\end{eqnarray}
The magnetic helicity spectrum is defined as the trace of the cross spectrum matrix,
\begin{eqnarray}
&& H_\mathrm{m}(\vec{k}) \equiv H_{ii}(\vec{k}) \\
&&   =-\frac{\mathrm{i}}{k^2}\left[k_1(S_{23}-S_{32})+k_2(S_{31}-S_{13}) + k_3(S_{12}-S_{21})\right] \nonumber.
\end{eqnarray}
Since $S_{23} = S_{32}^*$, $S_{23}-S_{32} = S_{23}-S_{23}^* = 2\mathrm{i} \mathrm{Im} S_{23}$,
\begin{eqnarray}
  H_\mathrm{m}(\vec{k}) &=& -\frac{\mathrm{i}}{k^2}\left(k_1 2\mathrm{i} \mathrm{Im} S_{23} + k_2 2\mathrm{i} \mathrm{Im}S_{31} + k_3 2\mathrm{i} \mathrm{Im}S_{12}\right) \nonumber \\
                        &=& \frac{2}{k^2}\left(k_1 \mathrm{Im} S_{23}+k_2 \mathrm{Im}S_{31}+k_3 \mathrm{Im}S_{12}\right),
\end{eqnarray}
where $\mathrm{Im}$ denotes the imaginary part of a complex number. 
The normalized magnetic helicity spectrum is then
\begin{eqnarray}
  \sigma_\mathrm{m}(\vec{k}) &=& \frac{\vec{k}H_m(\vec{k})}{S_{11}+S_{22}+S_{33}}  \nonumber \\
  &=& \frac{2\left(k_1 \mathrm{Im} S_{23}+k_2 \mathrm{Im}S_{31}+k_3 \mathrm{Im}S_{12}\right)}{k(S_{11}+S_{22}+S_{33})} \nonumber \\
      &=& \frac{2\left(V_{x0}\mathrm{Im}S_{yz} + V_{y0}\mathrm{Im}S_{zx} + V_{z0}\mathrm{Im}S_{xy}\right)}{V_{0}\left(S_{xx}+S_{yy}+S_{zz}\right)}. \label{eq:sigmamk}
\end{eqnarray}
Here, according to Taylor's hypothesis, the wave vector $\vec{k}$ is assumed to be aligned with the solar wind flow $\vec{V}_0$ in the spacecraft frame. We note that $V_{x0}$, $V_{y0}$, and $V_{z0}$ are three components of the solar wind flow $\vec{V}_0$.
Following \citet{Horbury2008}, the scale and time dependent mean magnetic field and flow velocity can be calculated using the envelope of the wavelet function.
Finally, the scale and time dependent magnetic helicity is
\begin{equation}\label{eq:sigmamst}
  \sigma_\mathrm{m}(s, t) = \frac{2[V_{x0}\mathrm{Im}(\tilde{B}_y^*\tilde{B}_z) + V_{y0}\mathrm{Im}(\tilde{B}_z^*\tilde{B}_x) + V_{z0}\mathrm{Im}(\tilde{B}_x^*\tilde{B}_y)]}{V_0 (|\tilde{B}_x|^2 + |\tilde{B}_y|^2 + |\tilde{B}_z|^2)}.
\end{equation}
Under normal solar wind conditions near 1 AU, the tangential flow $V_{y0}$ or $V_{z0}$ is usually very small compared to the radial component $V_{x0}$, so the expression is reduced to the first term only in the numerator \citep{Matthaeus1982}.
However, if the tangential flow becomes comparable to the radial flow, as may be the case near the Sun \citep{Kasper2019}, Equations \eqref{eq:sigmamk} and \eqref{eq:sigmamst} should be used.

\subsection{Cross helicity and residual energy}

Following \cite{Zhao2020ApJS}, we evaluated the normalized cross helicity and residual energy to measure the Alfv\'enicity of the structures.
The normalized cross helicity $\sigma_\mathrm{c}$ and residual energy $\sigma_\mathrm{r}$ were calculated from the Els\"asser variables $\vec{z}^\pm= \delta\vec{u} \pm \delta\vec{b}$ with $\delta\vec{b} = \delta\vec{B}/\sqrt{4\pi n_\mathrm{p} m_\mathrm{p}}$, $\delta\vec{u}$ is the fluctuating velocity field, $\delta\vec{B}$ is the fluctuating magnetic field, $n_\mathrm{p}$ is the proton number density, and $m_\mathrm{p}$ is the proton mass \citep[e.g.,][]{Zank2012}:
\begin{equation}\label{sigmac1}
  \sigma_\mathrm{c} = \frac{\langle z^{+2} \rangle - \langle z^{-2}\rangle}{\langle z^{+2} \rangle + \langle z^{-2}\rangle} = \frac{2 \langle \delta\vec{u} \cdot \delta\vec{b}\rangle}{\langle \delta u^2 \rangle + \langle \delta b^2 \rangle},
\end{equation}
and
\begin{equation}\label{sigmad1}
  \sigma_\mathrm{r} = \frac{2 \langle \vec{z}^{+} \cdot \vec{z}^- \rangle}{\langle z^{+2} \rangle  + \langle z^{-2} \rangle} = \frac{\langle \delta u^2 \rangle - \langle \delta b^2 \rangle}{\langle \delta u^2 \rangle + \langle \delta {b}^2 \rangle},
\end{equation}
where $\vec{z}^+$ ($\vec{z}^-$) represents the forward (backward) propagating modes with respect to the mean magnetic field orientation, and $\langle{z^+}^2\rangle$ and $\langle{z^-}^2\rangle$ represent the energy density in forward and backward propagating modes, respectively \citep{Zhao2020ApJS}.
Alfv\'enic fluctuations are associated with a high cross helicity (|$\sigma_\mathrm{c}| \sim 1$) and a low residual energy ($\sigma_\mathrm{r} \sim 0$). A high cross helicity indicates dominant energy in $\vec{z}^+$ or $\vec{z}^-$ modes, while a low residual energy indicates equipartition between kinetic and magnetic energies. These properties are characteristic of Alfv\'en waves. On the other hand, SFRs are not dominated by unidirectional Alfv\'enic waves, and magnetic energy usually dominates in SFRs compared to kinetic energy. The magnetic energy and kinetic energy defined here refer to the energy of the fluctuating magnetic field ($\delta \vec{B}$) and velocity ($\delta \vec{u}$), which do not include the mean magnetic field or mean flow velocity, and they are thus different from the magnetic and kinetic energies of large-scale ICMEs as determined in the common sense. As a result, SFRs typically have a low cross helicity (|$\sigma_\mathrm{c}| \sim 0$) and a highly negative residual energy ($\sigma_\mathrm{r} < 0$).  
In the following analysis, we use the wavelet analysis technique with a Morlet wavelet function \citep{Torrence1998} to construct spectrograms of the normalized magnetic helicity $\sigma_\mathrm{m}$, normalized cross helicity $\sigma_\mathrm{c}$, and normalized residual energy $\sigma_\mathrm{r}$.

\section{Magnetic flux ropes in the third and fourth \textit{PSP} encounters}\label{sec:results}
\subsection{Overview of PSP observations}
\begin{figure}[htbp]
\centering
\includegraphics[width=1.0\linewidth]{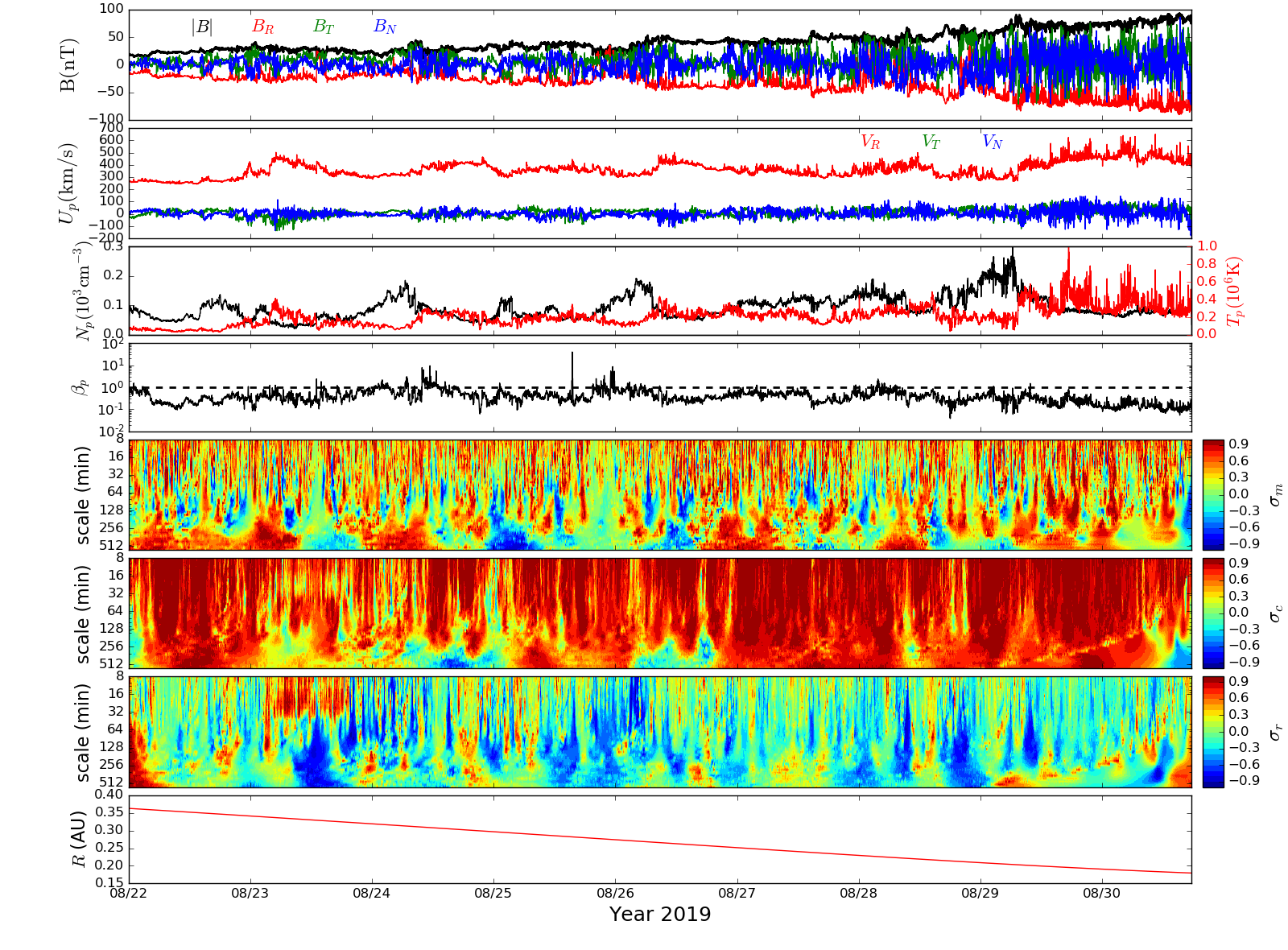} 
\caption{PSP in situ observations from its third inbound traverse from 2019 August 22 to 2019 August 30. The panels from top to bottom show the magnetic field magnitude ($|B|$) and three components in an RTN coordinate system; the solar wind velocity vector components; the proton number density ($N_\mathrm{p}$) and proton temperature ($T_\mathrm{p}$); the proton plasma beta ($\beta_\mathrm{p}$); the spectrograms of the normalized magnetic helicity ($\sigma_\mathrm{m}$); the normalized cross helicity ($\sigma_\mathrm{c}$); and the normalized residual energy ($\sigma_\mathrm{r}$). The bottom panel shows the radial distance of PSP.}\label{fig:3rd-full}
\end{figure}

Unless otherwise specified, magnetic field data from PSP/FIELDS \citep{Bale2016} and plasma data from the PSP/SWEAP/SPC \citep{Kasper2016} instruments are used in this work.
Figure\ \ref{fig:3rd-full} shows a nine-day plot of the PSP in situ magnetic field and plasma measurements during its third inbound traverse from 2019 August 22 to 2019 August 30. 
The panels show the magnetic field magnitude ($|B|$) and three components ($B_R$, $B_T$, and $B_N$); the solar wind speed components ($V_R$, $V_T$, and $V_N$); the proton number density ($N_\mathrm{p}$) and temperature ($T_\mathrm{p}$); the proton plasma beta ($\beta_\mathrm{p}$); the spectrogram of normalized magnetic helicity ($\sigma_\mathrm{m}$); the normalized cross helicity ($\sigma_\mathrm{c}$); the normalized residual energy ($\sigma_\mathrm{r}$); and the radial distance of PSP.
The closest distance to the Sun in the third orbit is about 0.17 AU on 2019 September 1. However, there are no plasma measurements near the third perihelion, and plasma data for the third outbound trajectory are not available until 2019 September 18, as the SPC instrument was powered off during this period due to an anomaly.
Therefore, we only show the results at radial distances down to $\sim 0.18$ AU, which are similar to the first two orbits.
In plotting the spectrograms of the normalized magnetic helicity $\sigma_m$, the normalized cross helicity $\sigma_c$, and the normalized residual energy $\sigma_r$, we used a one-day moving average to calculate the mean magnetic field and the mean flow speed. The normalized cross helicity is predominantly positive during this period, suggesting that unidirectional waves propagate in the direction opposite of the mean magnetic field.
Since the $B_R$ component is mostly negative during this nine-day plot, the positive cross helicity is consistent with outward propagating waves.
The normalized residual energy is close to zero in general, indicating an approximate equipartition between the magnetic and kinetic fluctuation energy. We also note that there are many patches of negative residual energy. The normalized magnetic helicity $\sigma_\mathrm{m}$ shows both positive and negative values, and there are regions of high $|\sigma_\mathrm{m}|$ with low $|\sigma_\mathrm{c}|$ and negative $\sigma_\mathrm{r}$. However, since this period is dominated by Alfv\'enic fluctuations ($\sigma_\mathrm{c}$ is close to 1 most of the time), only a small number of magnetic flux ropes are detected.

\begin{figure}[htbp]
\centering
\includegraphics[width=1.0\linewidth]{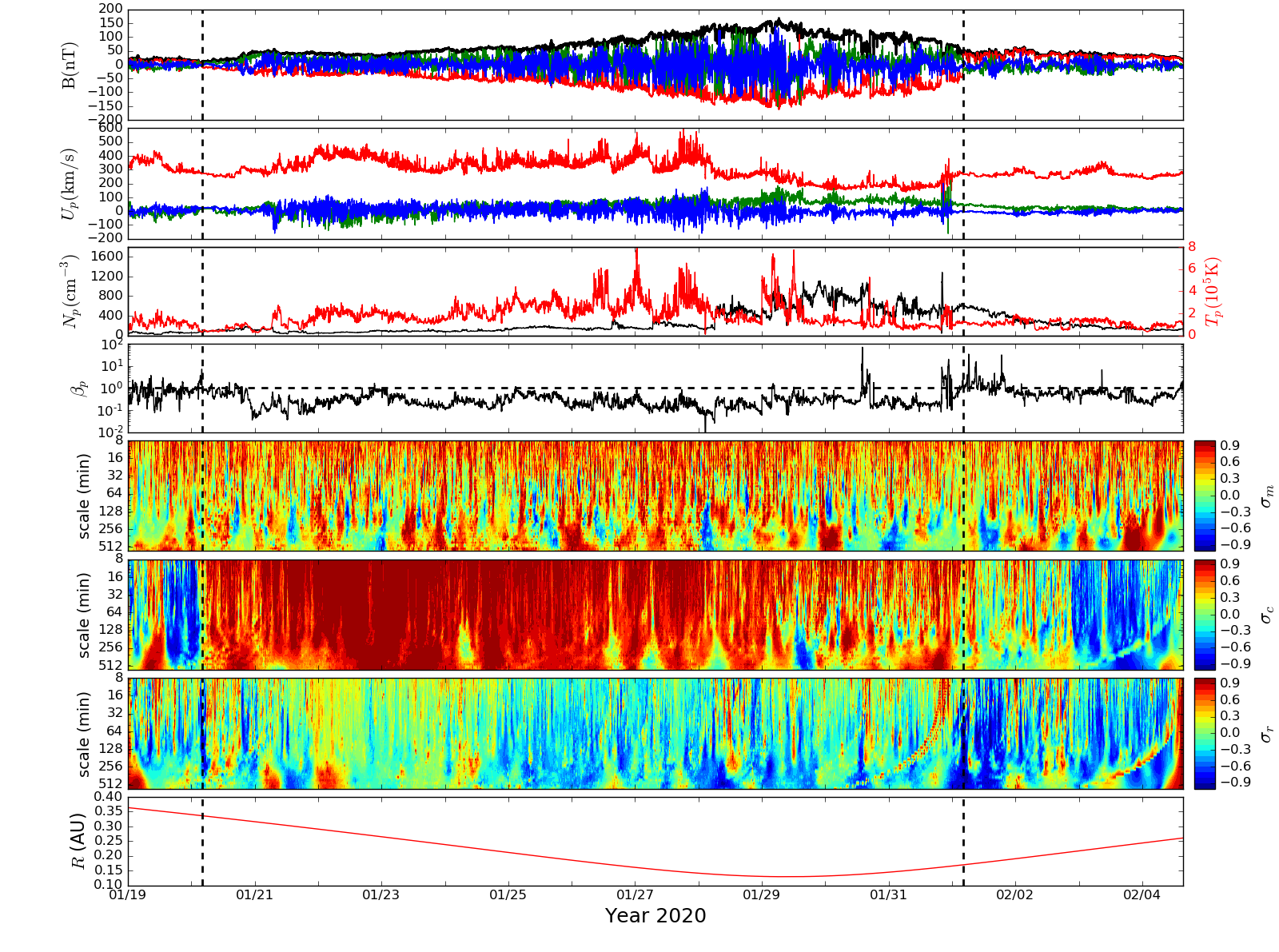} 
\caption{Same as Fig.\ \ref{fig:3rd-full}, but for the fourth orbit of PSP. Two HCS crossings are identified by the vertical dashed lines in each panel.}\label{fig:4th-full}
\end{figure}

Figure\ \ref{fig:4th-full} shows the magnetic field and plasma measurements during the fourth orbit of \textit{PSP} from 2020 January 19 to 2020 February 4. We note that, since the solar wind is not in the field of view of SWEAP/SPC during the period between 2020 January 29 and 2020 January 31, plasma data from the SWEAP/SPAN instrument is analyzed during these three days.
During the fourth orbit, \textit{PSP} is much closer to the Sun, and its fourth perihelion is at around 0.13 AU on 2020 January 29. It is worth noting that \textit{PSP} appears to sample slow solar wind near the fourth perihelion, starting on $\sim$\ 2020 January 28, which is indicated by the decreased proton bulk velocity and increased proton number density. Meanwhile, the normalized residual energy $\sigma_r$ starts to become negative as the normalized cross helicity $\sigma_c$ decreases from close to 1.
Another interesting feature of the fourth orbit is that there are two apparent crossings of the heliospheric current sheet (HCS). The first HCS crossing occurred on 2020 January 20, and the second was around 2020 February 1. The HCS crossings are characterized by sharp changes in the direction of the magnetic field radial component.
Accompanying the reversal of the magnetic field are changes of sign in normalized cross helicity $\sigma_\mathrm{c}$, as shown clearly in the spectrogram.
This is because outward propagating waves dominate on both sides of the HCSs.
After the second HCS crossing near 2020 February 1, the absolute value of the normalized cross helicity decreased to almost zero for the following two days before being dominated by negative values. This is due to\textit{PSP} sampling slow solar wind which was not dominated by Alfv\'enic fluctuations.

\begin{figure}[htbp]
\centering
\includegraphics[width=0.5\linewidth]{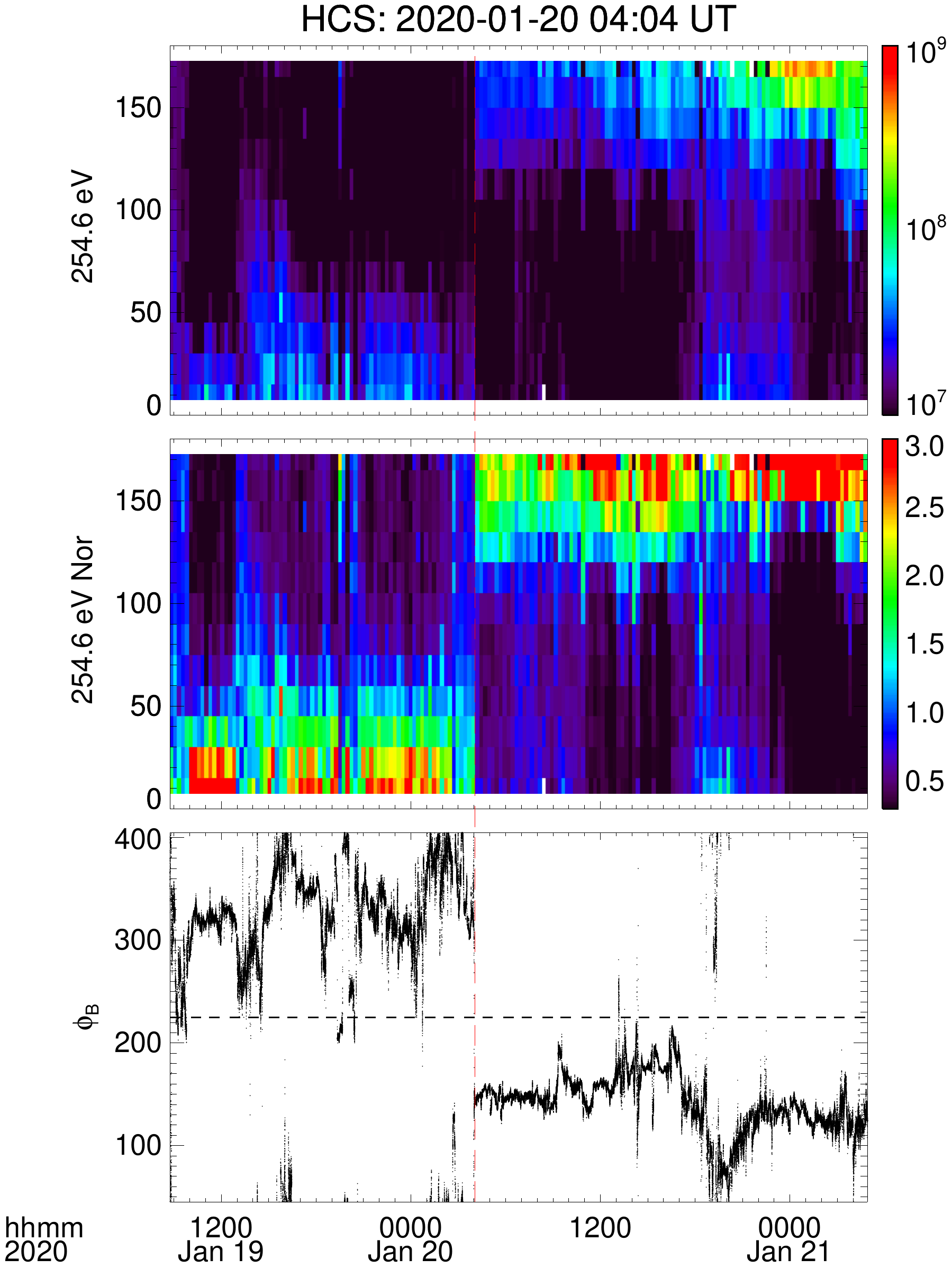}%
\includegraphics[width=0.5\linewidth]{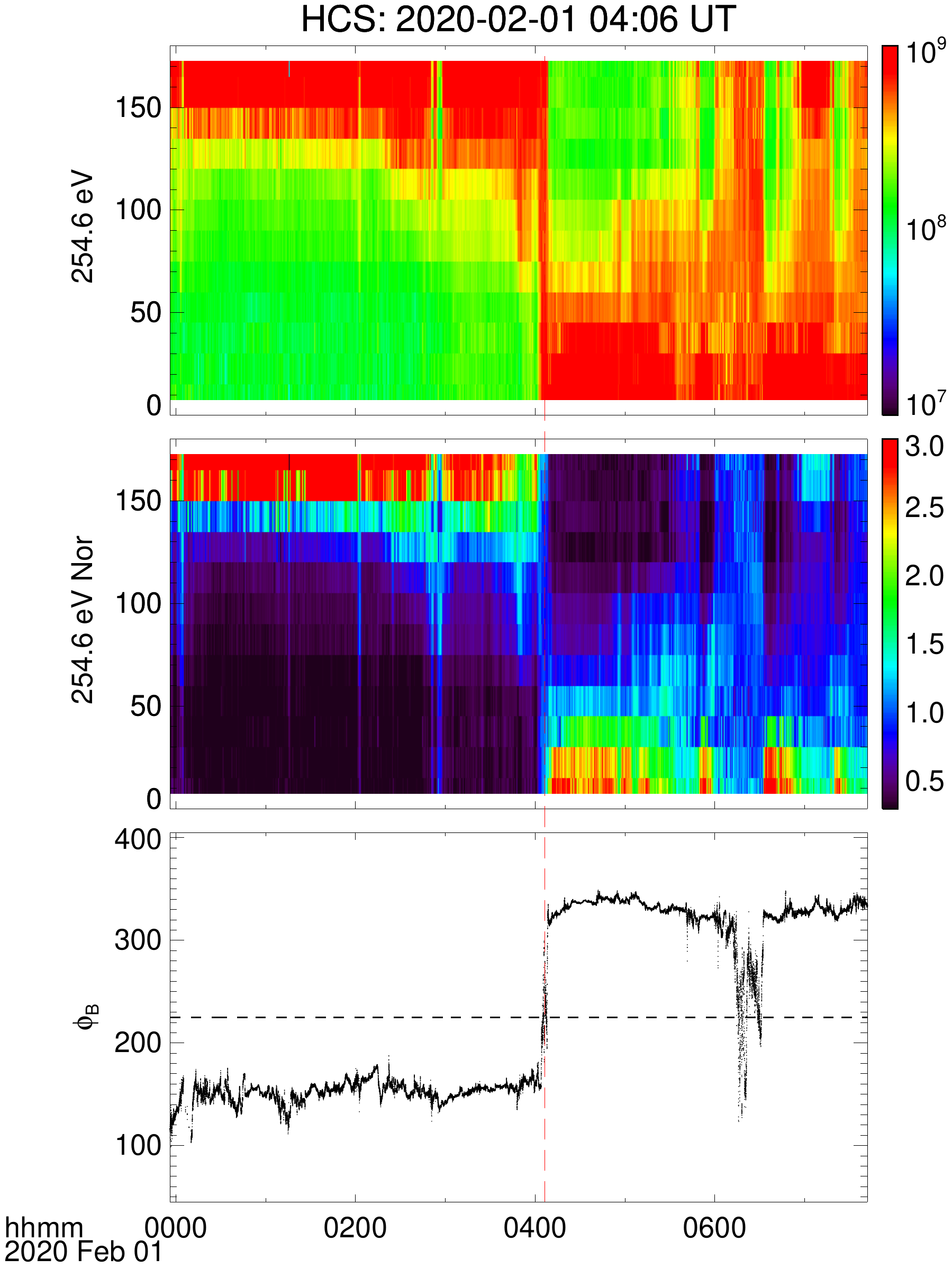}
\caption{Close-up view of the two HCS crossings. The top row plots the pitch angle distribution of superthermal electrons near the two HCS crossings and the second row shows the normalized distribution. The bottom panels plot the azimuthal angle of the magnetic field.} \label{fig:HCS}
\end{figure}

The two HCS crossings are further illustrated in Fig.\ \ref{fig:HCS}, where
the pitch angle distribution of 254.6 eV superthermal electrons and the normalized distribution can be seen. The normalization was derived by dividing the differential energy flux in each pitch angle by the mean value of all pitch angles during the same time period. Both crossings are clearly accompanied by a reversal in the electron strahl propagating direction. Furthermore, the azimuthal angle of the magnetic field changes by $\sim 180\degr$ at the HCS crossings.

During the second HCS crossing, the heliospheric plasma sheet (HPS) was observed around 2020 January 31, 22:00 UT -- February 1, 20:12 UT.
The HPS is characterized by an increase in proton plasma beta $\beta_\mathrm{p}$. The HPS in which the HCS is embedded is often thought to represent an extension of the streamer belt \citep[e.g.,][]{Liu2014}. 
As suggested by previous studies \citep{Cartwright2010, Khabarova2015, Adhikari2019}, small-scale magnetic flux ropes are most frequently observed near the HCS and HPS.
\textit{PSP} observations of the HCS and HPS make it possible to verify these findings at smaller radial distances.

\subsection{Identifying magnetic flux ropes}

Following \cite{Zhao2020ApJS}, we set a search criteria of $|\sigma_\mathrm{m}| \ge 0.7$ for magnetic flux ropes and removed structures that were too small (scale less than 5 minutes) or too large (scale higher than 300 minutes). Structures that have very small scales may be contaminated by discontinuities and are not the focus of this study. The very large structures may not be reliable since they usually fall outside of the cone of influence of the wavelet spectra \citep{Torrence1998}. Structures that further satisfy the condition $|\sigma_\mathrm{c}| \le 0.3$ and $\sigma_\mathrm{r} \le -0.5$ are identified as magnetic flux ropes.
These conditions are set to exclude most Alfv\'en waves or Alfv\'enic structures as they typically have high values of normalized cross helicity $|\sigma_\mathrm{c}|$ and low values of normalized residual energy $\sigma_\mathrm{r}$.
We find a total of 715 structures with $|\sigma_\mathrm{m}| \ge 0.7$ during the 21-day period of the third orbit of PSP, 21 of which are identified as magnetic flux ropes. During the 17-day period of PSP's fourth orbit, we find 840 structures with a high magnetic helicity, among which 34 magnetic flux ropes are identified.
The numbers may be compared with the results from \citet{Zhao2020ApJS}, where a total of 1253 structures with an enhanced magnetic helicity were identified during a 31-day period of the first orbit of PSP, including 40 magnetic flux ropes.
The occurrence rate of magnetic flux ropes during the fourth orbit of \textit{PSP} is much higher than its previous orbits. This could be interpreted in two ways. First, turbulence may be more effective in generating SFRs in the pristine solar wind, and these structures, whose lifetime is short, are not observed at larger distances. A second interpretation is that some structures are connected to the coronal magnetic field and thus can only be observed near the Sun. 

\begin{figure}[htbp]
\centering
\includegraphics[width=1.0\linewidth]{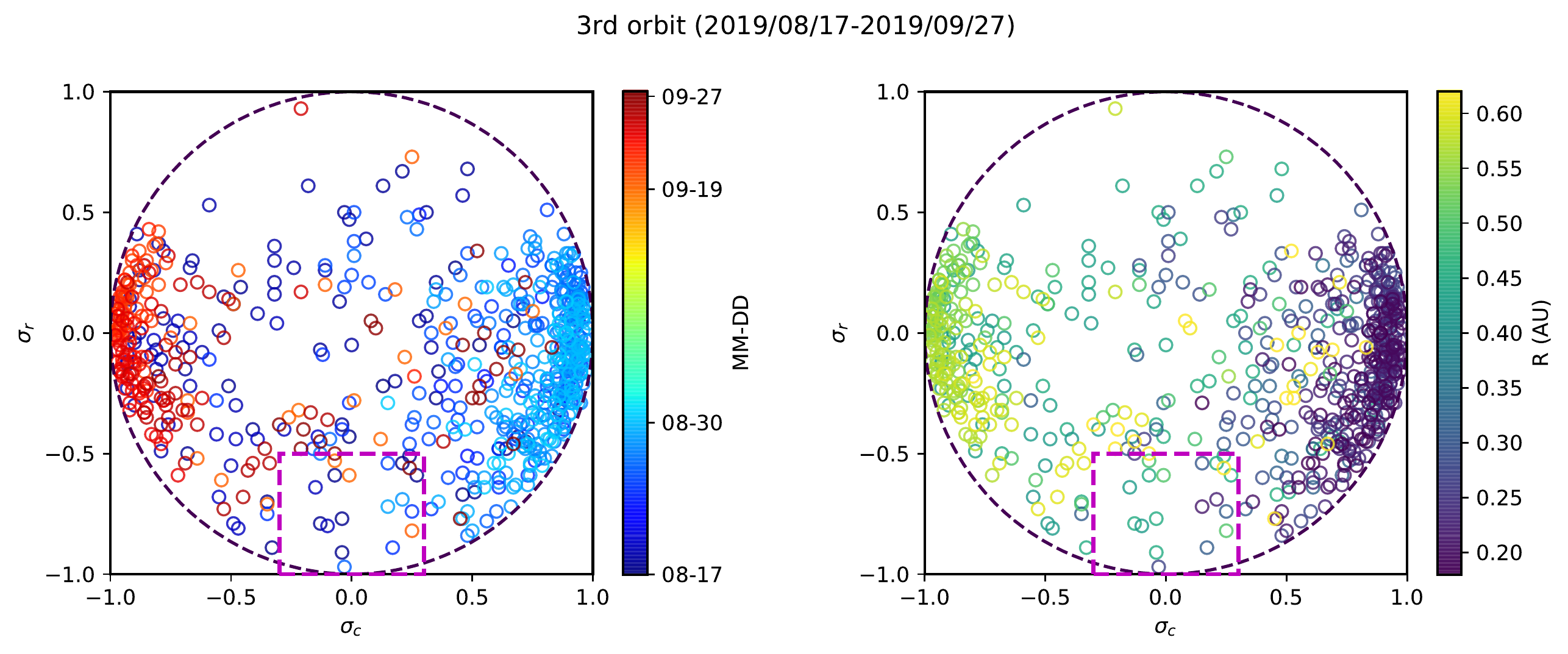}
\caption{Normalized cross helicity $\sigma_\mathrm{c}$ vs. normalized residual energy $\sigma_\mathrm{r}$ for structures with a high magnetic helicity ($|\sigma_m| \ge 0.7$) identified in the third orbit. We note that there is a lack of plasma data during the periods from 2019 August 31 to 2019 September 18, as well as between 2019 August 20 and 2019 August 21. In the left panel, the scatter circles are colored according to the date. In the right panel, the points are colored according to the radial distance to the Sun. The dashed-dotted circle represents $\sigma_\mathrm{c}^2+\sigma_\mathrm{r}^2=1$. The magenta rectangular box represents the region that likely contains flux rope structures with a low value of normalized cross helicity $|\sigma_c| \le 0.3$ and a highly negative residual energy 
$\sigma_r \le -0.5$.} \label{fig:o3}
\end{figure}

\begin{figure}[htbp]
\centering
\includegraphics[width=1.0\linewidth]{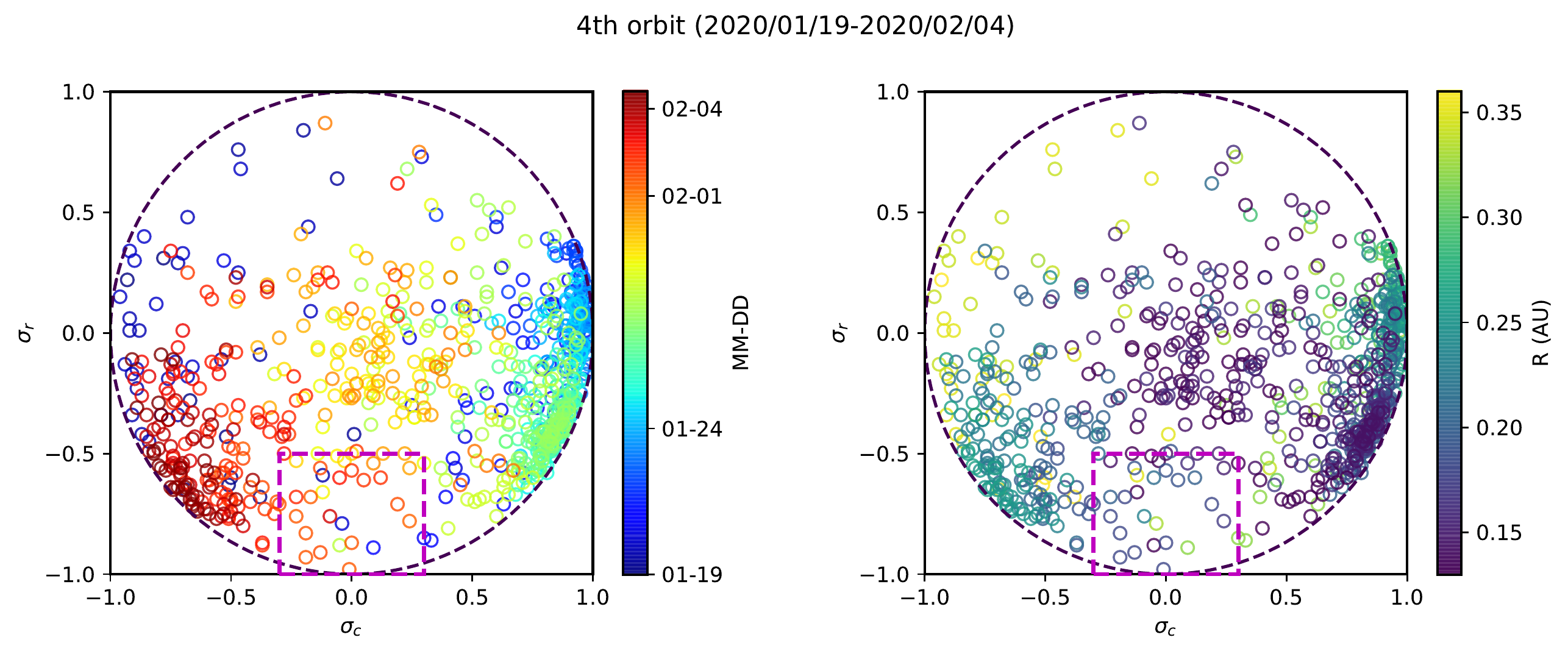}
\caption{Normalized cross helicity $\sigma_\mathrm{c}$ vs. normalized residual energy $\sigma_\mathrm{r}$ for structures with $|\sigma_m|\ge 0.7$ in the fourth orbit. The format is the same as in Fig.\ \ref{fig:o3}.} \label{fig:o4}
\end{figure}

Figure\ \ref{fig:o3} displays the normalized cross helicity $\sigma_\mathrm{c}$ versus the normalized residual energy $\sigma_\mathrm{r}$ for the identified structures with $|\sigma_\mathrm{m}| \ge 0.7$ during the 21-day course of the third orbit (2019/08/17-2019/08/19, 2019/08/22-2019/08/30, and 2019/09/19-2019/09/27).
We note that there is a lack of plasma data during the periods from 2019 August 31 to 2019 September 18, as well as between 2019 August 20 and 2019 August 21.
Magnetic flux ropes that satisfy our selection criteria are within the magenta rectangular box ($|\sigma_\mathrm{c}| \le 0.3$ and $\sigma_\mathrm{r} \le -0.5$).
Figure\ \ref{fig:o4} shows structures with a high magnetic helicity during the 17-day period from 2020 January 19 to 2020 February 4 in the fourth orbit.
In both orbits, structures are predominantly Alfv\'enic, as indicated by the relatively large number of points with $|\sigma_c| >0.5$. There are significantly more structures with low values of $\sigma_c$ during the fourth orbit and they are typically dominated by magnetic fluctuations with $\sigma_\mathrm{r} < 0$.
From the left panel of Fig.\ \ref{fig:o4}, we find that the most likely flux rope structures (inside the magenta rectangular box) that were observed in the fourth orbit are near 2020 February 1, which corresponds to the second HCS crossing, thus indicating that SFRs are frequently generated via magnetic reconnection across the HCS.
The right panel of Fig.\ \ref{fig:o4} shows that these structures are relatively close to the Sun, although some other flux ropes are observed farther away from the Sun.
In general, for pure Alfv\'en waves, the sum in quadrature of the normalized cross-helicity and the normalized residual energy should be 1 (i.e., $\sigma_\mathrm{c}^2+\sigma_\mathrm{r}^2 =1$ in Fig.\ \ref{fig:o3} and Fig.\ \ref{fig:o4}). As is shown in Fig.\ \ref{fig:o3}, in the third orbit, most of the scatters lie near the circle of radius 1, with few scatters in the middle of it. In contrast, during the fourth orbit, the scatters are more uniformly distributed, with relatively more scatters distributed in the middle of the circle with a radius of 1. It indicates that as the Sun is approached more and more closely, the solar wind becomes more and more populated by structures, including both propagating Alfv\'enic fluctuations and advected structures. 

\begin{figure}[htbp]
\centering
\includegraphics[width=0.5\linewidth]{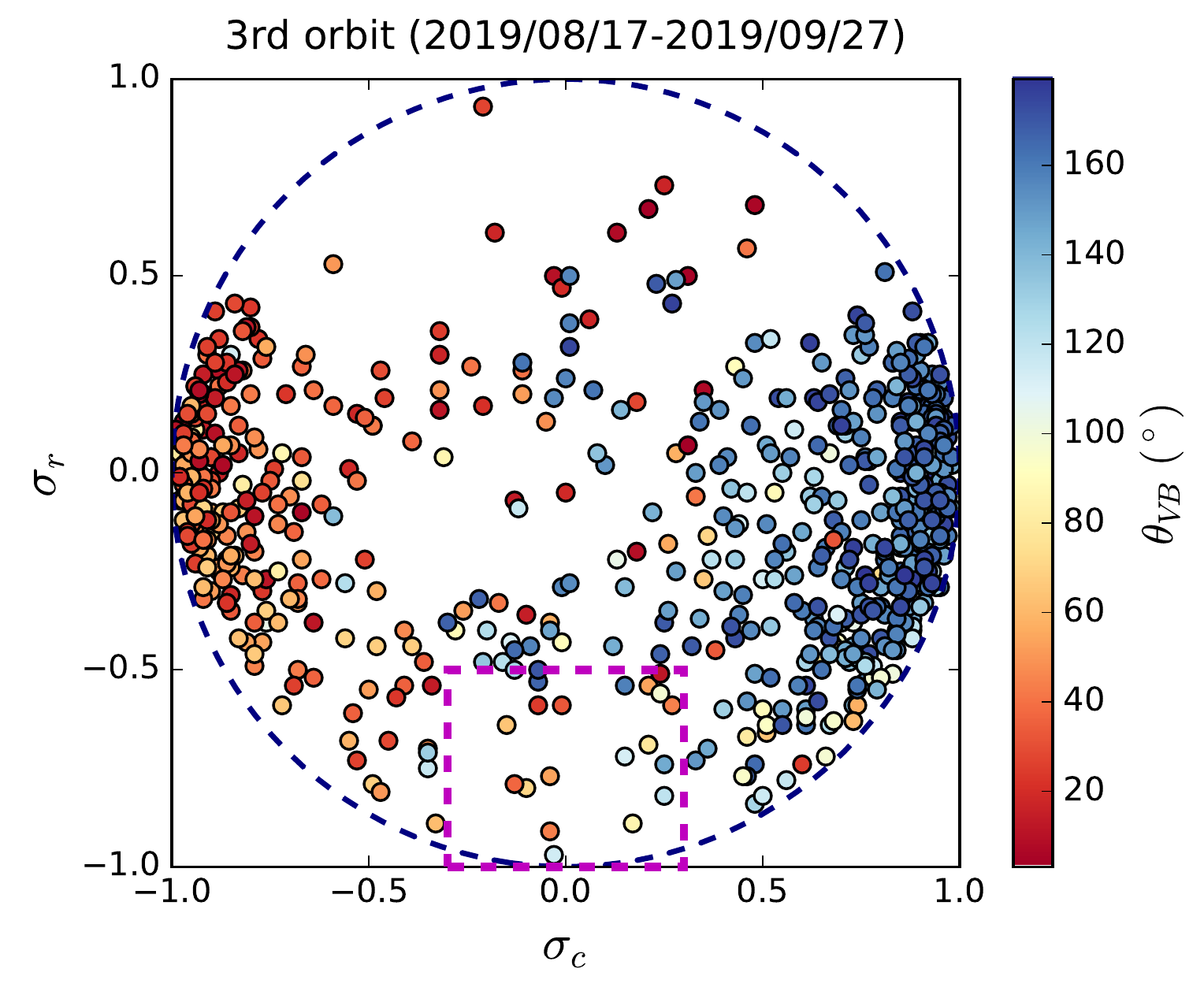}%
\includegraphics[width=0.5\linewidth]{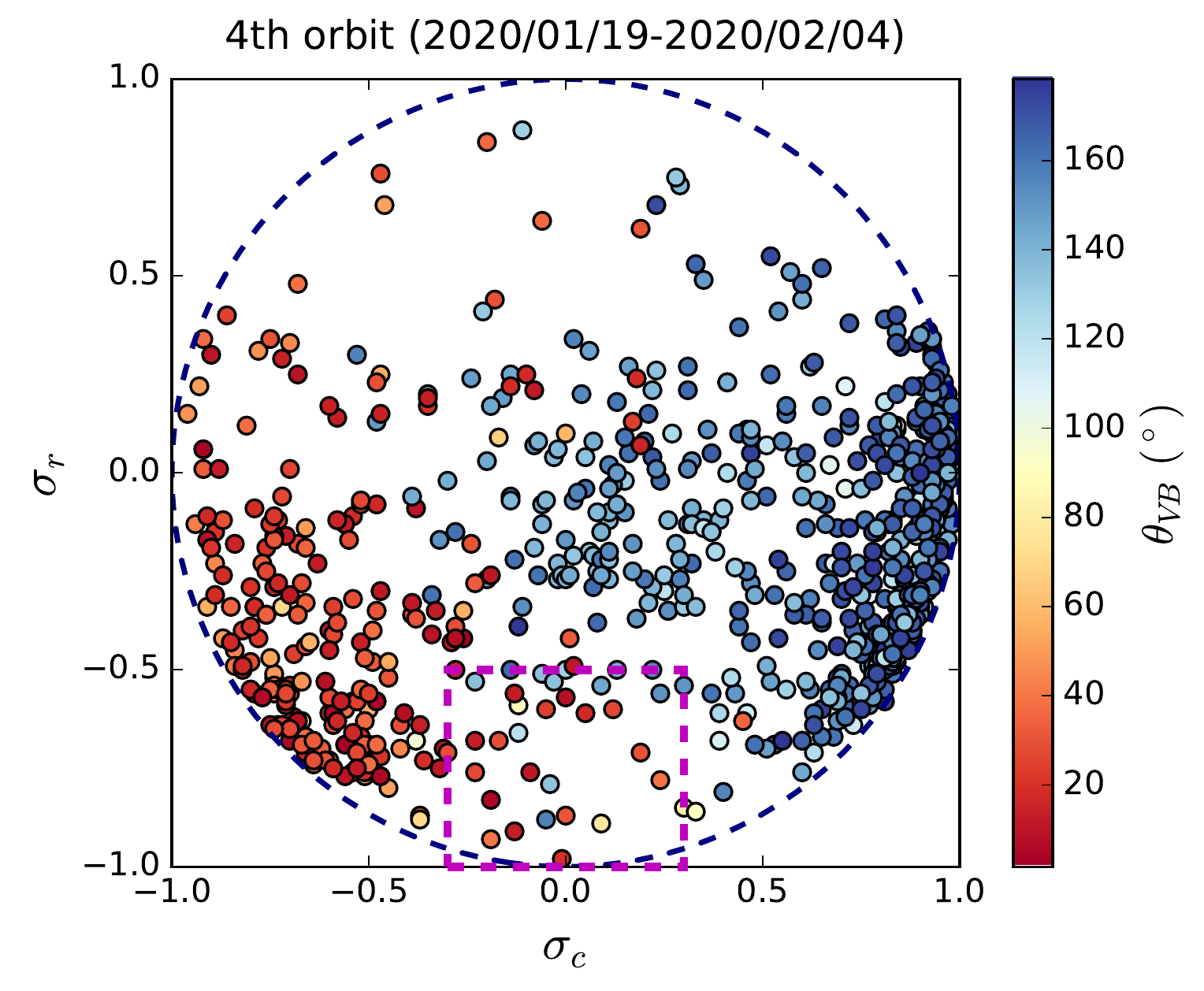}
\caption{Normalized cross helicity $\sigma_\mathrm{c}$ vs. normalized residual energy $\sigma_\mathrm{r}$ for structures with $|\sigma_m|\ge 0.7$ in the third (left panel) and fourth (right) orbits. The scatters are color coded by the angle between the mean magnetic field and the mean flow speed.} \label{fig:theta}
\end{figure}

As \textit{PSP} continues to approach the Sun, it tends to sample field-aligned flows as one would expect from the spiral interplanetary magnetic field model of \citet{Parker1958}. To assess the possible selection bias resulting from the field alignment of solar wind flow, we calculated the angle between the mean magnetic field and the mean solar wind flow directions for all of the identified structures, as is shown in Fig.\ \ref{fig:theta}.
Similar to Fig.\ \ref{fig:o3} and \ref{fig:o4}, we plotted the normalized cross helicity $\sigma_\mathrm{c}$ and normalized residual energy $\sigma_\mathrm{r}$ for structures with a high magnetic helicity.
The results from the third and fourth orbits are plotted on the left and right panels, respectively. 
The figure shows that the Alfv\'enic fluctuations are almost entirely observed in field-aligned or anti-field-aligned flows. The angle $\theta_{\mathrm{VB}}$ is close to $180\degr$ when $\sigma_\mathrm{c} \simeq 1$ and $\theta_{\mathrm{VB}} \simeq 0\degr$ when $\sigma_\mathrm{c} \simeq -1$. Inside the rectangular box, which likely contains magnetic flux ropes, there is a mixture of different angles ranging from $0\degr$ to $180\degr$. This suggests that flux ropes are almost universally present and they are seen as soon as the flow deviates from being highly aligned or anti-aligned.
Therefore, the turbulence may appear to be predominantly Alfv\'enic because \textit{PSP} samples mainly highly field-aligned flows, which inhibits the detection of flux ropes in large parts of the flow. 

\subsection{Occurrence rate and solar wind parameter dependence}

\begin{figure}[htbp]
\centering
\includegraphics[width=1.0\linewidth]{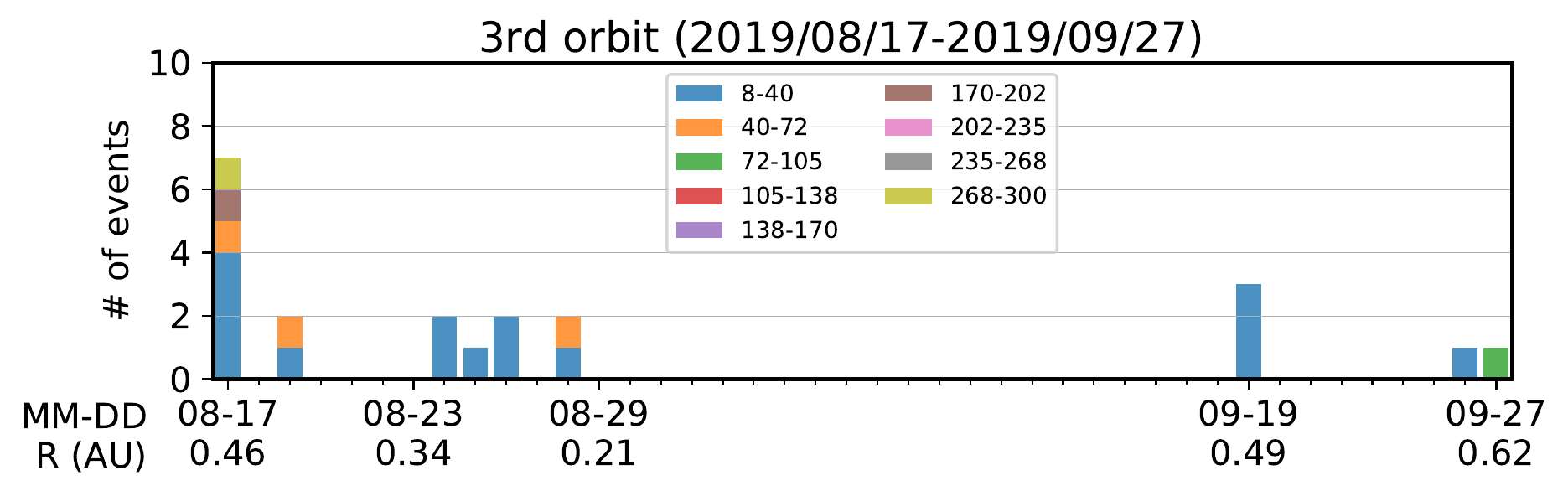}
\includegraphics[width=1.0\linewidth]{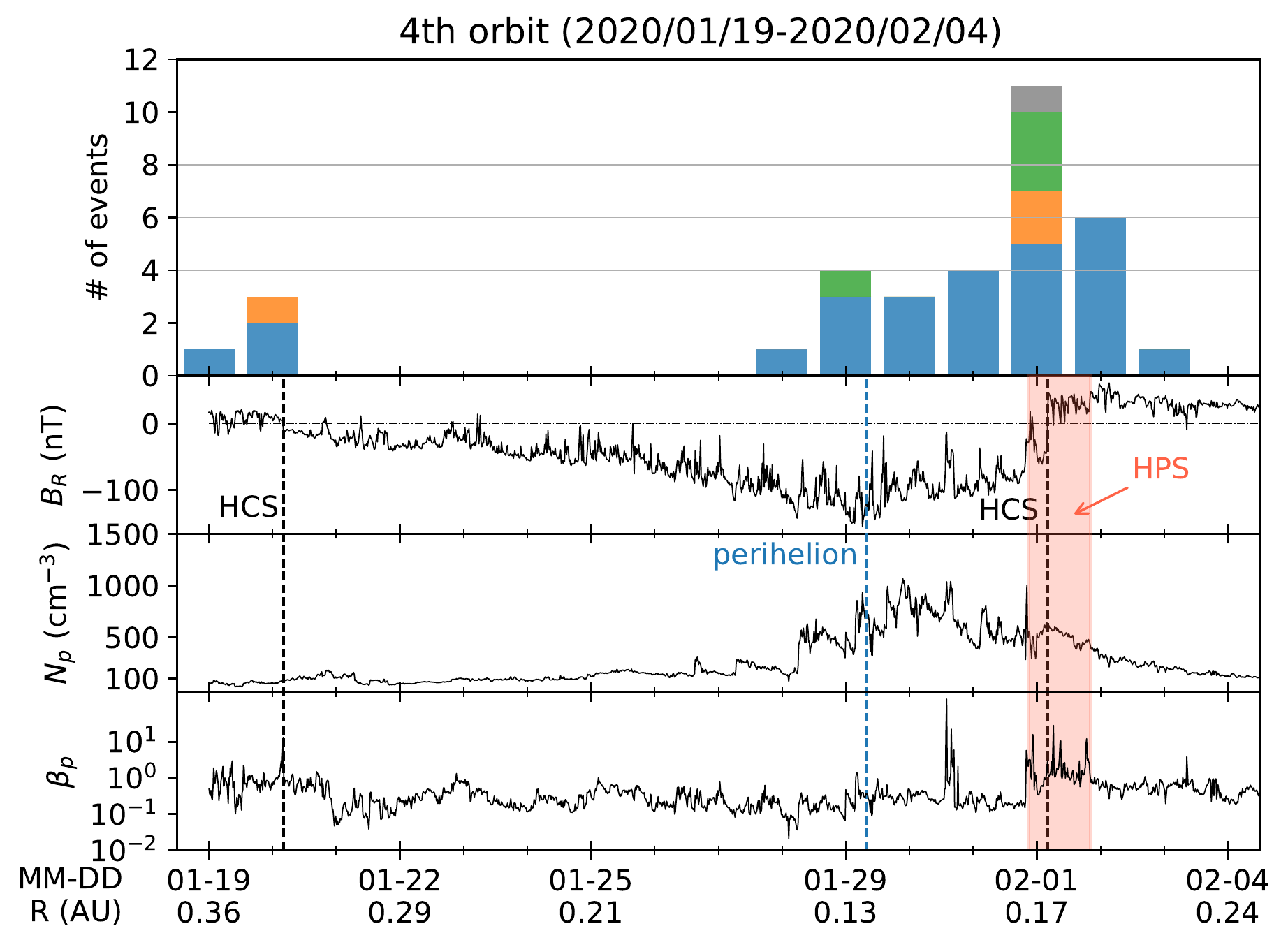}
        \caption{Counts of magnetic flux ropes per calendar day in the third (top panel) and fourth (bottom panels) orbits. Flux ropes with different durations (in minutes) are distinguished by different colors. For the fourth orbit, the radial magnetic field $B_R$, proton density $N_p$, and proton plasma beta $\beta_p$ are shown in the bottom three panels. The two HCS crossings are identified by black vertical dashed lines, the fourth perihelion ($\sim$0.13 AU) is indicated by a blue vertical dashed line, and the HPS is shown by the pink shaded area.} \label{fig:rate}
\end{figure}

Figure\ \ref{fig:rate} shows the occurrence rate of magnetic flux ropes using our magnetic helicity-based detection technique.
We counted the number of flux ropes in each calendar day and show the results as bar plots;
the third and fourth orbit results are also plotted in this figure. 
We distinguish flux ropes with different durations (in minutes) using different colors. Furthermore, the fourth orbit's 10 minutes moving-averaged radial magnetic field $B_R$, proton density $N_p$, and proton plasma beta $\beta_p$ are also shown in addition to two HCS crossings and the fourth perihelion as well as the HPS, which is considered as the extension of the streamer belt at a larger heliocentric distance. 
The figure shows that the most of the flux ropes have a short duration (less than 72 minutes). In comparing the results from the two orbits, one can see that the occurrence rate is higher in the fourth orbit than in the third orbit, which is likely due to the prevailing slow solar wind and the HCS crossing along with the fact that the \textit{PSP} sampled solar wind is closer to the Sun.
For the fourth orbit, it can be seen from the figure that most of the flux ropes are observed between 2020 January 29 (the fourth perihelion) and 2020 February 2. The dates correspond to the period when PSP is in the vicinity of the streamer belt and HCS crossing \citep[e.g.,][]{Hu2018}. Longer-duration flux ropes are also mostly observed near this period. We note that some of the long-duration flux ropes may originate from the streamer belt blobs that can be observed from coronagraph images, but the connection needs a more dedicated analysis in order to be verified.
No flux ropes are observed between 2020 January 21 and 2020 January 27, corresponding to a period of solar wind streams with high values of cross helicity (close to 1) and near zero residual energy, as can be seen from Fig.\ \ref{fig:4th-full}.
We note that the first HCS crossing on 2020 January 20 also appears to cause a slight enhancement in the occurrence rate of magnetic flux ropes, but the effect is much weaker compared to the second HCS crossing.
This is probably due to the arrival of relatively fast solar wind streams shortly after the first HCS crossing, while the second HCS crossing is in a very slow solar wind associated with the streamer stalk plasma.

\begin{figure}[!tp]
\centering
\includegraphics[width=1.0\linewidth]{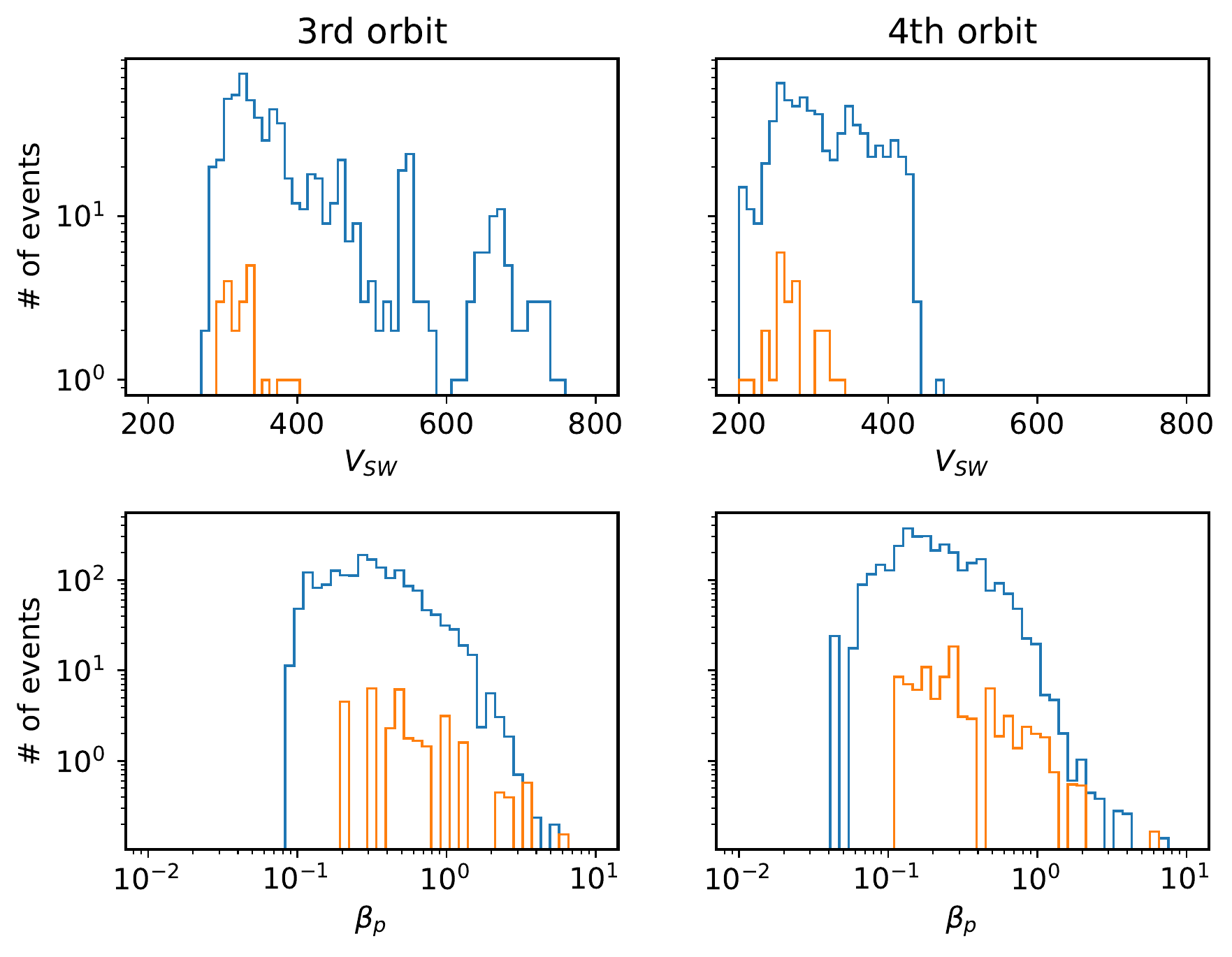}
\caption{Histograms of the solar wind speed and proton plasma beta for identified high magnetic helicity structures. The orange lines represent the distribution of magnetic flux ropes, and the blue lines represent other structures. The third and fourth orbits are shown in the left and right panels, respectively.} \label{fig:hist}
\end{figure}

Histograms of the solar wind velocity $V_{sw}$ and proton plasma beta $\beta_p$ of the high magnetic helicity structures are shown in Fig.\ \ref{fig:hist}. 
The distribution of magnetic flux ropes and other structures are shown in Fig.\ \ref{fig:o3} and Fig.\ \ref{fig:o4}. The third and fourth orbits are shown in the left and right panels, respectively.
From the figure, we clearly see that the solar wind speed reaches lower values (down to $\sim 200$ km/s) in the fourth orbit.
The plasma beta also reaches lower values in the fourth orbit, shown in the bottom row.

\begin{figure}[htbp]
\centering
\includegraphics[width=1.0\linewidth]{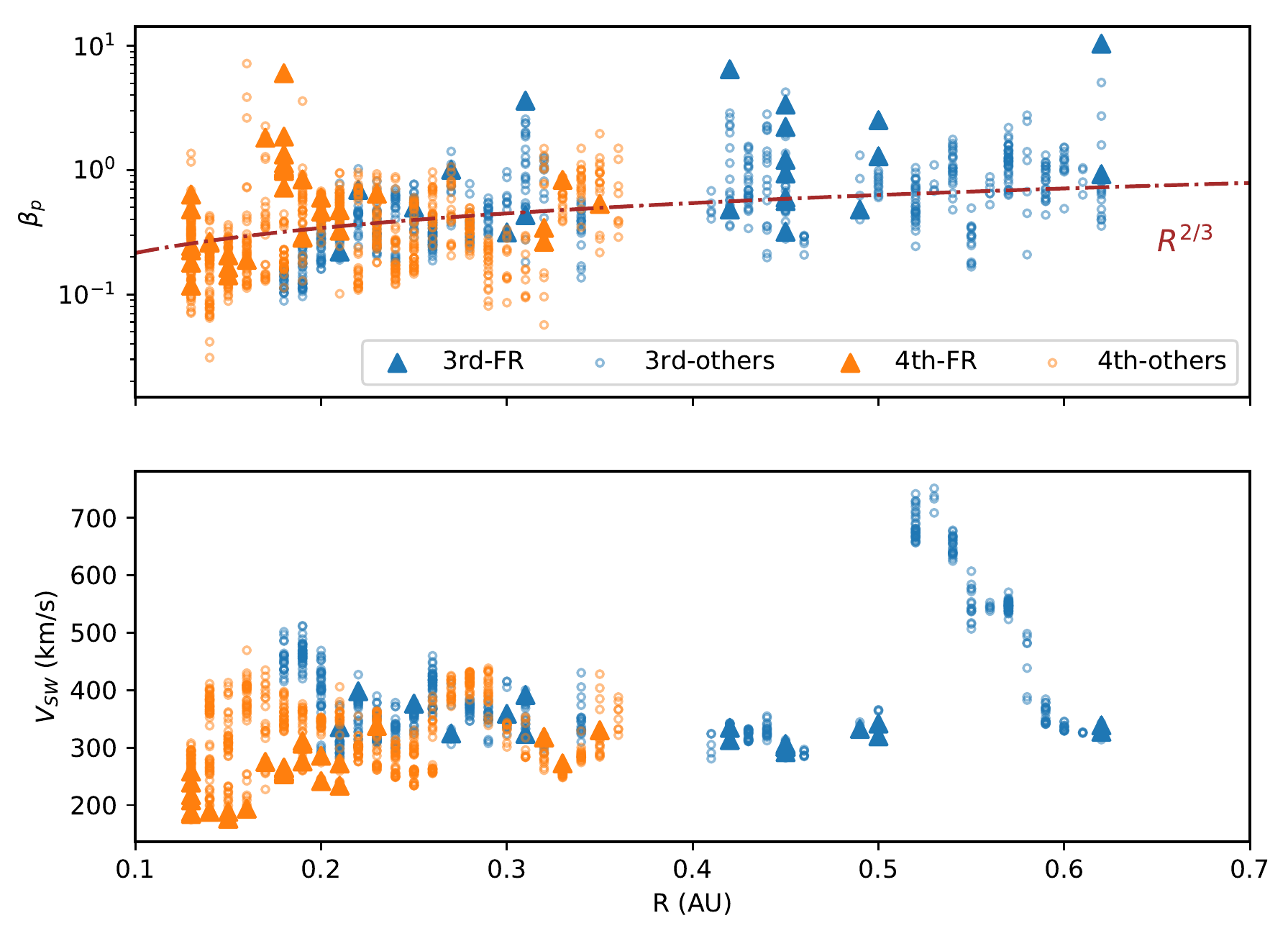}
\caption{Radial dependence of the proton plasma beta (top panel) and solar wind speed (bottom panel) of the identified high magnetic helicity structures in the third orbit (blue markers) and fourth orbit of \textit{PSP} (orange markers). Magnetic flux ropes are plotted as triangles and other high magnetic helicity structures are plotted as circles. The red dashed-dotted curve in the top panel represents the adiabatic prediction of the plasma beta in the solar wind with $\beta \propto R^{2/3}$.} \label{fig:radial}
\end{figure}

The lower values of the solar wind speed and proton plasma beta in the fourth orbit are probably due to the closer distance to the Sun.
Figure\ \ref{fig:radial} further shows the radial dependence of the proton plasma beta and solar wind speed in the top and bottom panels.
Magnetic flux ropes and other structures are also plotted.

As might be expected, both the plasma beta and solar wind speed show an increasing trend with radial distance.
Beta values smaller than $\sim 0.1$ are almost exclusively observed in the fourth orbit.
The red dashed-dotted line indicates the theoretical prediction of the plasma beta in the solar wind. Assuming an adiabatic equation of state for an ideal gas of pressure $P$ and density $\rho$, $P\rho^{-\gamma}=constant$ with a polytropic index $\gamma$, and a radial expansion and a radial magnetic field, $\rho \propto R^{-2}$ and $B\propto R^{-2}$, then the plasma beta has a radial dependence of $\beta \propto R^{4-2\gamma}$.
This is illustrated by the red dashed-dotted curve in the top panel, which shows a $R^{2/3}$ ($\gamma = 5/3$) radial dependence, predicted by the adiabatic expansion of solar wind.
As one gets closer to the Sun, the observed plasma beta dependence departs slightly from the adiabatic prediction. 
It is important to notice that if the polytropic index $\gamma < 5/3$, we obtain a radial evolution of $\beta \propto R^a$ where $a>2/3$. Such a curve would fit the data better. A smaller polytropic index ($\gamma<5/3$) corresponds to heating of the solar wind, which may be due to a 2D turbulence cascade \citep{Zank2017, Zank2018}. This is consistent with the view that some small-scale magnetic flux ropes are generated locally by turbulence \citep{Zank2017, Zheng2018}.
From the bottom panel, we note that the solar wind speed during the fourth orbit of \textit{PSP} is relatively slow ($< 500$ km/s) compared to its third orbit. At the fourth perihelion, the solar wind speed is around 200 km/s, which is also shown in Fig.\ \ref{fig:4th-full} and Fig.\ \ref{fig:hist}. In summary, we find that the identified small magnetic flux ropes in the third and fourth orbits \textit{PSP} mostly lie in slow solar wind and exhibit a wide range of proton plasma beta values, although a low plasma beta is generally considered to be a very reliable signature of large-scale magnetic flux ropes, that is, CMEs or ICMEs. This is in nice agreement with previous statistical studies at 1 AU \citep[e.g.,][]{Yu2014}.

\subsection{Flux ropes near the HCS}
\begin{figure}[htbp]
\centering
\includegraphics[width=1.0\linewidth]{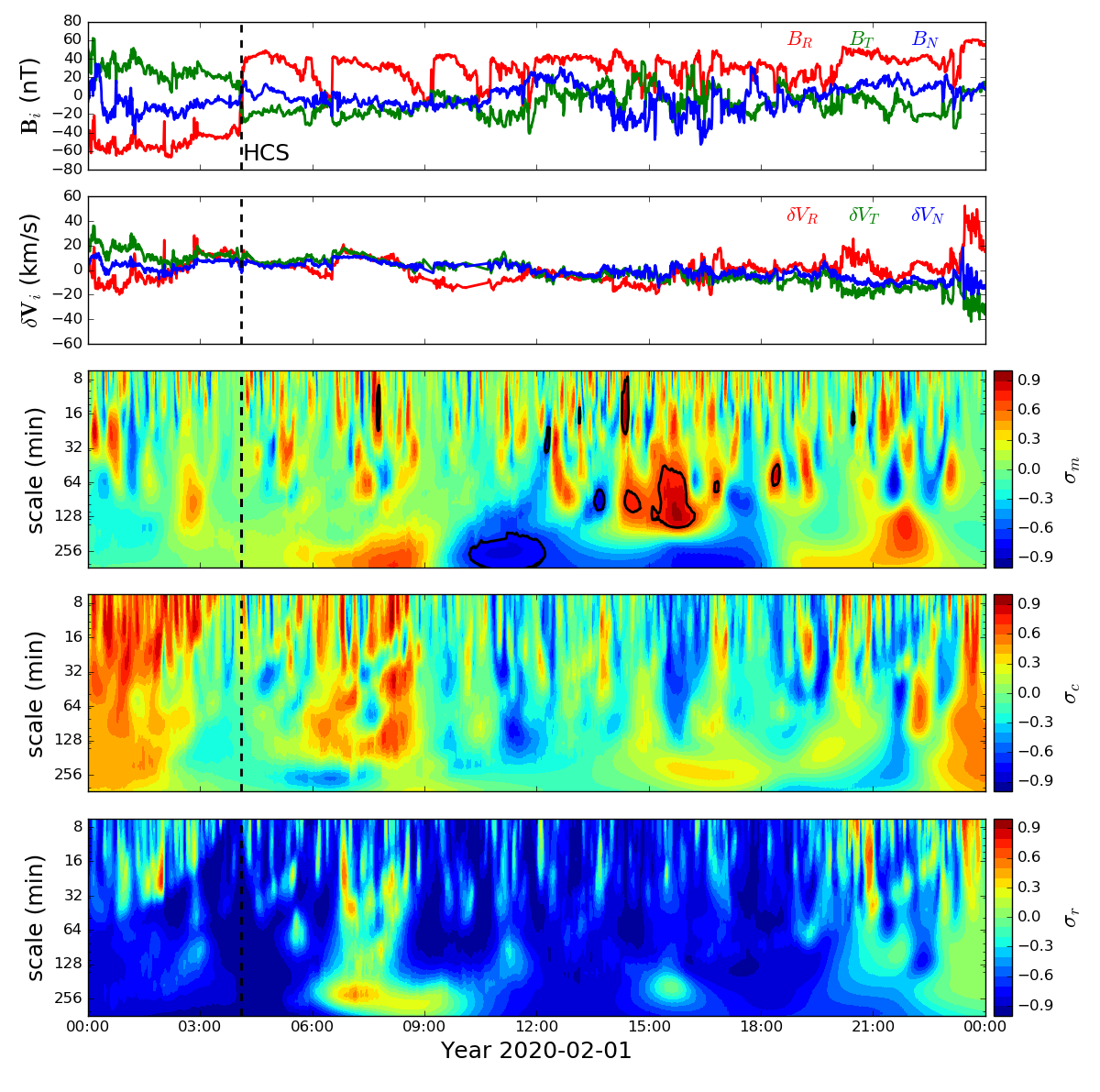}
\caption{Expanded view of the second HCS crossing on 2020 February 1. The top two panels show the magnetic field and fluctuating velocity components. The bottom three panels show the wavelet spectrograms of the normalized magnetic helicity $\sigma_\mathrm{m}$, the normalized cross helicity $\sigma_\mathrm{c}$, and the normalized residual energy $\sigma_\mathrm{r}$. Magnetic flux ropes that satisfy our selection criteria ($|\sigma_m| \ge 0.7$ and $|\sigma_c| \le 0.3$ and $\sigma_r \le -0.5$) are labeled by contour lines in the $\sigma_\mathrm{m}$ plot.} \label{fig:crossing}
\end{figure}

As discussed above, magnetic flux ropes are observed more frequently near the second HCS crossing. Here, we show in Fig.\ \ref{fig:crossing} an expanded view of this period on 2020 February 1.
The top two panels show the magnetic field and fluctuating velocity components.
The bottom three panels show the wavelet spectrograms of the normalized magnetic helicity $\sigma_\mathrm{m}$, the normalized cross helicity $\sigma_\mathrm{c}$, and the normalized residual energy $\sigma_\mathrm{r}$.
Magnetic flux ropes that satisfy our selection criteria ($|\sigma_m| \ge 0.7$ and $|\sigma_c| \le 0.3$ and $\sigma_r \le -0.5$) are labeled in the $\sigma_\mathrm{m}$ plot.
Figure\ \ref{fig:crossing} shows that flux ropes with both positive and negative $\sigma_\mathrm{m}$ are observed. From the sign of $\sigma_\mathrm{m}$, we can determine the chirality of underlying fluctuations at a specific scale. Positive $\sigma_\mathrm{m}$ corresponds to right-handed chirality and a negative value corresponds to left-handed chirality.

The largest flux rope shown in the figure has a characteristic timescale of more than 200 minutes, while small structures are on the order of 10 minutes.
Rotation of the magnetic field can be seen clearly in some of the large-scale structures.
We note that some structures have an overlapping time range.
From a turbulence perspective, this can be understood as the coexistence of structures at different scales as a result of the turbulent cascade.
\begin{table*}[htbp]
\caption{List of identified magnetic flux ropes on 2020 February 01 near the second HCS crossing.} \label{tab:list}
\centering
\begin{tabular}{cccccccc}
\hline\hline
Central time & Scale & $\left<\sigma_\mathrm{m}\right>$ & $\left<\sigma_\mathrm{c}\right>$ & $\left<\sigma_\mathrm{r}\right>$ & $R$ (AU) & $\left<V_{\mathrm{sw}}\right>$ & $\left<\beta_\mathrm{p}\right>$ \\
 (UT) & (minutes) & & & & & (km/s) & \\
\hline
07:46:04 & 14   & 0.77 & 0.24  & -0.78 & 0.17 & 275 & 1.82 \\
11:14:44 & 262  & -0.76  & -0.17 & -0.68 & 0.18 & 258 & 1.86 \\
12:17:19 & 28   & -0.76  & -0.23 & -0.76 & 0.18 & 263 & 1.06 \\
13:09:00 & 17   & 0.75  & -0.01  & -0.98 & 0.18 & 259 & 1.04  \\
13:40:51 & 92 & -0.78 & -0.19 & -0.83 & 0.18 & 258 & 1.13 \\
14:22:30 & 13 & 0.82 & -0.19 & -0.93 & 0.18 & 256 & 0.73 \\
14:33:11 & 95 & 0.74 & 0.00 & -0.87 & 0.18 & 254 & 0.98 \\
15:39:48 & 90 & 0.8 & -0.3 & -0.71 & 0.18 & 259 & 1.34 \\
16:48:31 & 71 & 0.72 & 0.19 & -0.71 & 0.18 & 265 & 1.02 \\
18:24:02 & 55 & 0.77 & -0.13 & -0.91 & 0.18 & 265 & 6.03 \\
20:27:52 & 18 & -0.76 & -0.13 & -0.56 & 0.19 & 276 & 0.84 \\
\hline
\end{tabular}
\end{table*}

As an example, 
a list of the identified flux ropes near the second HCS crossing on 2020 February 1 is provided in Table \ref{tab:list}.\ It contains the following information:
the central time of the structure; the characteristic timescale of the structure in minutes; the average $\sigma_\mathrm{m}$, $\sigma_\mathrm{c}$, and $\sigma_\mathrm{r}$; the radial distance from the Sun in AU; the average solar wind speed within the structure in km/s; and the average plasma beta within the structure.

\section{An estimate of error}\label{sec:error}

\begin{figure}[htbp]
\centering
\includegraphics[width=0.5\linewidth]{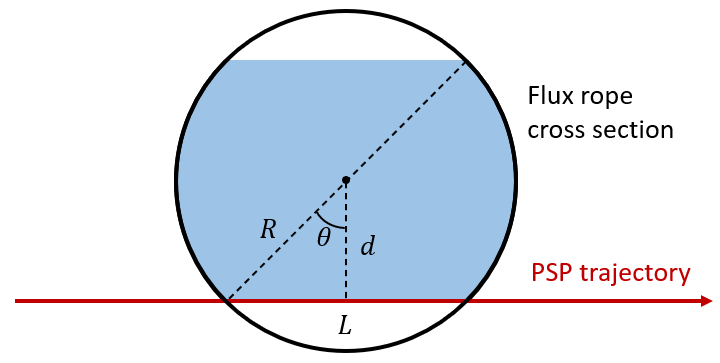}
\caption{Illustration of the error estimation. See text for details.} \label{fig:error}
\end{figure}

Since magnetic flux ropes are 3D structures in nature, any observations from a single vantage point inevitably have errors and uncertainties when identifying flux ropes.
The technique that we used is no different in this regard.
In particular, the detection and analysis of a magnetic flux rope are more reliable if the spacecraft crosses the center of the structure and less reliable if the spacecraft crosses the structure at its flank.
This has recently been studied in detail by \cite{Telloni2020} using simulated spacecraft trajectories.
The idea is illustrated in Fig.\ \ref{fig:error}.

Here, we assume that the flux rope has a circular cross section of radius $R$ and the spacecraft (\textit{PSP}) crosses the flux rope at a distance $d$ from the center of the structure.
We let $L$ be the length traversed by \textit{PSP}, so that the fraction of the flux rope sampled by \textit{PSP} is $L (2R)^{-1}$, or expressed in terms of the angle $\theta$ as $\sin\theta = L (2R)^{-1}$.
\cite{Telloni2020} show that the inaccuracy of the magnetic helicity based analysis is related to the above fraction.
Specifically, the error is small when the fraction is larger than $\sim$50\% and the method becomes unreliable when the fraction is smaller than $\sim$50\%. For example, an error of $\sim$40\% in the measured flux rope time scale can be expected when the fraction is 50\%.
As a result, we may use $L (2R)^{-1} \ge 0.5$ as the condition for the method to be reliable, or $d/R = \cos\theta \le \sqrt{3}/2$.
If we suppose that magnetic flux ropes propagate approximately isotropically (near the ecliptic plane where \textit{PSP} is), then it can be estimated that the probability of the method being reliable is $\sim \sqrt{3}/{2} \simeq 87\%$.
In Fig.\ \ref{fig:error}, assuming that the red trajectory is the threshold, then the method is reliable when the PSP trajectory falls in the blue shaded region.

The above error estimate considers only a specific source of error. In reality, there are certainly other sources of errors and uncertainties.
For example, the cross section of the flux rope is probably not a perfect circle; the magnetic field lines may not have an idealized helical structure as assumed by \cite{Telloni2020}; the flux rope axis may not be normal to the spacecraft trajectory; and the propagation direction of flux ropes may not be isotropic.
However, these sources of uncertainties are not easy to quantify.
Nevertheless, our simple analysis does suggest that more than 80\% of the results from our technique are likely to be reasonably accurate.
We note that our selection criterion of $|\sigma_\mathrm{m}| \ge 0.7$ excludes some events where the distance between the spacecraft path and flux rope center is large. To a certain extent, using a higher threshold for magnetic helicity would have excluded more cases with large errors.

\section{Summary and discussions}\label{sec:summary}

We have applied a wavelet analysis to determine the magnetic helicity, cross helicity, and residual energy in a systematic search for magnetic flux rope structures during the third and fourth orbits of PSP\ around the Sun.
The analysis technique was developed by \cite{Zhao2020ApJS} and applied to the first orbit of \textit{PSP}.
The calculation of normalized, reduced magnetic helicity has been improved here to incorporate a finite rotational flow, which may be significant near the Sun \citep{Kasper2019}.
As in \citet{Zhao2020ApJS}, structures with a high normalized reduced magnetic helicity ($|\sigma_\mathrm{m}| \ge 0.7$) were first identified, which may include both magnetic flux ropes and Alfv\'en waves.
Magnetic flux ropes were further selected as structures with low normalized cross helicity ($|\sigma_\mathrm{c}| \le 0.3$) and highly negative normalized residual energy ($\sigma_\mathrm{r} \le -0.5$).

To summarize, we draw the following conclusions.
\begin{enumerate}
\item We find a total of 715 (840) high magnetic helicity structures in the third (fourth) orbit, of which 21 (34) are classified as magnetic flux ropes.
The occurrence rate for all of those high magnetic helicity structures is $\sim$ 34 per day ($\sim$ 49 per day) for the third (fourth) orbit, compared with $\sim$ 40 per day during the first orbit. For flux ropes, the occurrence rate is $\sim$ 1 per day ($\sim$ 2 per day) for the third (fourth) orbit, compared with $\sim$ 1 per day during the first orbit. The fourth orbit has a higher occurrence rate of magnetic flux rope structures compared to previous orbits.
\item The solar wind speed achieves much lower values in the fourth orbit due to \textit{PSP} reaching a closer radial distance to the Sun. The solar wind speed is $\sim 200$ km/s near the fourth perihelion. There are some high plasma beta regions in the outbound leg of the fourth orbit, which are thought to be related to the crossings of the HPS and HCS.
\item Magnetic flux ropes are more likely to be observed in the slow solar wind, while fast solar wind is dominated by Alfv\'enic structures.
This is consistent with our results from the first orbit of \textit{PSP} \citep{Zhao2020ApJS} and is in good agreement with previous statistical studies at 1 AU \citep{Yu2014}.
In particular, \textit{PSP} observed a dynamic streamer belt during the outbound trajectory of the fourth orbit, where the slow solar wind is thought to originate.
We find a concentration of magnetic flux ropes with a wide range of duration in the vicinity of the observed streamer stalk region.
\item PSP observed two HCS crossings during its fourth orbit. The region near the HCS crossing shows an obvious increase in the counts of small magnetic flux ropes, especially for the second crossing when PSP was embedded in the HPS, where the solar wind is rather slow.
\item A simple estimation of error based on the results of \cite{Telloni2020} suggests that more than 80\% of the structures are likely to be accurately calculated, although there are other sources of uncertainties that have not been quantified.
\end{enumerate}

In conclusion, our study identifies magnetic flux ropes in a new regime closer to the Sun as measured during the third and fourth orbits of \textit{PSP}. Small-scale magnetic flux ropes have been observed throughout the heliosphere. The presence of those coherent structures is closely related to the nature of solar wind turbulence \citep{Zank2017, Zank2020} and possibly the energization of charged particles \citep{Zank2014, Zhao2019ApJa, Adhikari2019}.
The high occurrence rate and significance of magnetic flux ropes near the streamer stalk and the HCS crossing is important for understanding the turbulent dynamics of this region.

\begin{acknowledgements}
We acknowledge the partial support of the NSF EPSCoR RII-Track-1 Cooperative Agreement OIA-1655280, and partial support from a NASA Parker Solar Probe contract SV4-84017. 
Parker Solar Probe was designed, built, and is now operated by the Johns Hopkins Applied Physics Laboratory as part of NASA’s Living with a Star (LWS) program (contract NNN06AA01C). Support from the LWS management and technical team has played a critical role in the success of the Parker Solar Probe mission.  
\end{acknowledgements}

\bibpunct{(}{)}{;}{a}{}{,} 
\bibliographystyle{aa}
\bibliography{psprobe}
\end{document}